# Fano resonance enabled infrared nano-imaging of local strain in bilayer graphene

Jing Du(杜靖)[1,2,†], Bosai Lyu(吕博赛)[1,2,†], Wanfei Shan(单琬斐)[1,2,†], Jiajun Chen(陈佳俊)[1,2], Xianliang Zhou(周先亮)[1,2], Jingxu Xie(谢京旭)[3], Aolin Deng(邓奥林)[1,2], Cheng Hu(胡成)[1,2], Qi Liang(梁齐)[1,2], Guibai Xie(谢贵柏)[4], Xiaojun Li(李小军)[4], Weidong Luo(罗卫东)[1,2,5,*], Zhiwen Shi(史志文)[1,2,*]

[1]Key Laboratory of Artificial Structures and Quantum Control (Ministry of Education), Shenyang National Laboratory for Materials Science, School of Physics and Astronomy, Shanghai Jiao Tong University, Shanghai, China.

[2]Collaborative Innovation Center of Advanced Microstructures, Nanjing, China.

[3]Institute of Physics, Xi'an Jiao Tong University, Xi'an, China.

[4]National Key Laboratory of Science and Technology on Space Science, China Academy of Space Technology (Xi'an), Xi'an, China.

[5]Institute of Natural Sciences, Shanghai Jiao Tong University, Shanghai, China.

[†] These authors contribute equally to this work.

[*] To whom correspondence should be addressed. Email: zwshi@sjtu.edu.cn, wdluo@sjtu.edu.cn

**Abstract:**

Detection of local strain at the nanometer scale with high sensitivity remains challenging. Here we report near-field infrared nano-imaging of local strains in bilayer graphene through probing strain-induced shifts of phonon frequency. As a non-polar crystal, intrinsic bilayer graphene possesses little infrared response at its transverse optical (TO) phonon frequency. The reported optical detection of local strain is enabled by applying a vertical electrical field that breaks the symmetry of the two graphene layers and introduces finite electrical dipole moment to graphene phonon. The activated phonon further interacts with continuum electronic transitions, and generates a strong Fano resonance. The resulted Fano resonance features a very sharp near-field infrared scattering peak, which leads to an extraordinary sensitivity of ~0.002% for the strain detection. Our studies demonstrate the first nano-scale near-field Fano resonance, provide a new way to probe local strains with high sensitivity in non-polar crystals, and open exciting possibilities for studying strain-induced rich phenomena.

**Main text:**

Strain plays an important role in condensed matter physics and materials science because it can strongly modify the mechanical, electrical, and optical properties of a material and even induce structural phase transitions.[1-3] Strain effects are extremely inspiring in atomically thin 2D materials because they are fundamentally capable of sustaining much larger strain compared to their bulk counterpart. Various strain-induced phenomena in 2D materials, including modulation of electrical properties,[1,4] transitions from an indirect-to-direct bandgap,[2] valley polarization,[5] soliton-dependent plasmon reflection,[6] structural phase transitions,[7] pseudo magnetic gauge field,[3] and brightening of dark excitons[8] have been investigated. To fully understand these strain-induced phenomena, it is critical to be aware of the strain distribution at the nanometer scale.

In order to probe the fine structure of local strains, one needs to employ strain-detection techniques with both high spatial resolution and high sensitivity. Classical methods like X-ray diffraction (XRD)[9] and neutron scattering[10] offer high sensitive measurements of strain, but they usually require bulk samples and their spatial resolution is poor. Transmission electron microscopy (TEM)[11] provides lattice scale imaging of local strain but it is not compatible with samples on substrates. Recent studies reveal that the scanning near-field optical microscopy (SNOM)[5,6,12,13] provides a powerful tool for probing local strains in polar crystal structures with both high spatial resolution and high sensitivity.[14-16] With the SNOM technique, the strain-induced shift of phonon frequency can be directly detected through resonant Rayleigh scattering, from which strain amplitudes can be deduced. Typically, a blue shift of phonon frequencies corresponds to compressive strain, and a red shift corresponds to tensile strain. Such detection necessarily requires the optical phonon of the target material to be infrared active, i.e., its lattice vibration carries electrical dipole moment that can interact efficiently with infrared light. For examples, optical phonons in polar crystals of $SiO_2$, $Al_2O_3$, hBN are infrared active, and therefore strains in these materials are detectable with SNOM; while optical phonons in nonpolar crystals, such as silicon, diamond, and graphene, are infrared inactive, and strains in them are normally undetectable with SNOM.

Nevertheless, bilayer graphene, a representative nonpolar 2D crystal, has recently attracted tremendous research interests owing to its unique physical properties. First, its band structure is tunable with an external vertical electrical field, and a large bandgap up to ~300 meV is available.[17-

[19] Second, bilayer graphene in certain magic angle twists features strongly correlated phenomena,[20-25] such as superconductivity,[21,23] Mott insulator,[20,22,24] and magnetism,[25] and has become an ideal platform for simulating the Hubbard model and understanding the mechanism of unconventional superconductivity. However, strain distribution and strain effect on the properties of bilayer graphene have so far remained unknown.

Previous far-field infrared spectroscopic studies have shown that infrared inactive phonon (Fig. 1a) in bilayer graphene can turn into infrared active by applying a vertical electrical field (Fig. 1b), because the applied vertical field will break the symmetry of the two layers and introduce electrical dipole moment through the electron-phonon coupling.[26-29] More interestingly, such electron-dressed phonon features a Fano resonance when interacting with infrared photons, and forms a tunable Fano system.[27,28] Enlightened by these studies, we combine Fano activation with infrared nano-spectroscopy, and report high-sensitive infrared nano-imaging and quantitative analysis of local strains in bilayer graphene. The developed strain detection method features spectacularly a high sensitivity of 0.002%. Such high sensitivity originates from Fano interference induced extremely sharp phonon peak. Furthermore, strain-induced shifts of phonon frequencies are calculated using first-principles density functional theory (DFT).

Bilayer graphene flakes are mechanically exfoliated onto $SiO_2$/Si substrates from bulk graphite. Wrinkles and cracks naturally exist in the exfoliated graphene samples due to the uncontrollable mechanical exfoliation process. Local strains are expected near the end of those wrinkles and cracks. The direct probe of local strain in bilayer graphene is performed under electrical gating through infrared nano-imaging via a home-built SNOM setup (see Methods for more details), the schematic of which is shown in Fig. 1d. In brief, an infrared laser beam is focused on the apex of a metal-coated AFM tip. Backscattered light from the tip apex is collected and measured by an MCT detector placed in the far field. Near-field scattering amplitude is extracted through a homodyne detection scheme. The tip-enhanced infrared scattering provides a local probe of the material infrared responses with ~20 nm spatial resolution. Lithographically fabricated Au/Ti electrode allows applying a gate voltage $V_g$ (Fig. 1c) when performing the SNOM measurement. The electrical gating can dope the graphene with charge carriers, and more importantly, provide a vertical electrical field that activates the graphene phonon.

Figure 1e presents the topography images of a bilayer graphene sample with wrinkles. Figure

1f-h present near-field IR amplitude images of the same sample at different excitation frequencies. IR image in Fig. 1f is taken at excitation of 1575 cm$^{-1}$, which is off the phonon resonance frequency 1585 cm$^{-1}$. The bright straight lines correspond to the wrinkles. The bright curved lines across the bilayer graphene sample are layer-stacking domain walls, which are absent in the topography image. Such detection of domain walls agrees with previous near-field optical studies.[5,6,30] IR image of Fig. 1g is taken at excitation of 1580 cm$^{-1}$, which is slightly lower than the phonon resonance frequency ~1585 cm$^{-1}$. In addition to wrinkles and domain walls, prominent dark spots are seen near the end of the wrinkles. Figure 1h is taken at excitation of 1588 cm$^{-1}$, which is slightly higher than the phonon frequency. Obvious bright spots show up at exactly the locations of dark spots in Fig. 1g. Presumably, the observed IR contrast spots in Fig. 1g and h near the end of wrinkles are induced by the local strains. Strain can alter the original state of chemical bonds by elongating or shortening the bonding length, which will shift resonating phonon frequency and modify the near-field optical response. Specifically, a compressive strain can shift the optical phonon frequency to a higher frequency, while a tensile strain will shift the optical phonon in bilayer graphene to a lower frequency.

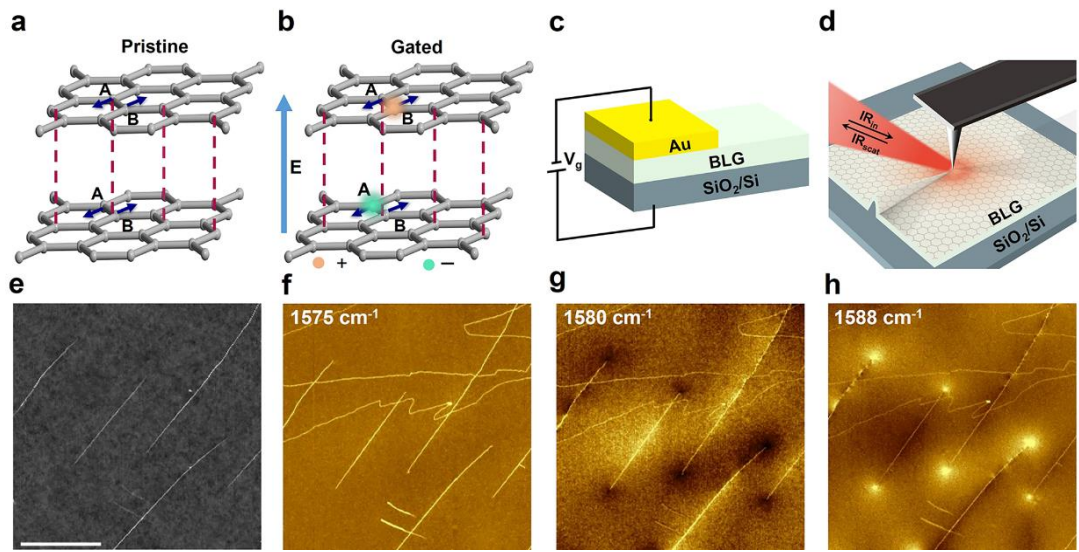

**Fig. 1. Near-field IR Nano-Imaging of local strain in exfoliated bilayer graphene.** (a) Symmetric optical phonon in a pristine Bernal stacked bilayer graphene with zero dipole moment. (b) Field-induced charge reapportion leads to infrared activity in the optical phonon. (c) Illustration of the bilayer graphene device. (d) Illustration of the near-field infrared nanoscopy measurement of local strain in bilayer graphene. Infrared light is focused on the apex of a metal-coated AFM tip, and the backscattered infrared light is collected by an MCT detector in the far field. (e) AFM topography image of an exfoliated bilayer graphene sample with wrinkles on a SiO$_2$/Si substrate. (f) Image of near-field infrared scattering amplitude of the same sample as in (e), taken at 1575 cm$^{-1}$, which is off the phonon resonance frequency. The bright straight-line features are wrinkles. The bright curve features across the bilayer graphene samples are domain walls. (g) Corresponding near-field infrared nanoscopy image taken at 1580 cm$^{-1}$, showing obvious dark spots at the end of wrinkles. (h) Corresponding near-field infrared nanoscopy taken at a frequency of 1588 cm$^{-1}$, showing prominent bright spots near the wrinkle ends. Scale bar 5 $\mu m$.

To further confirm the measured IR contrast is induced by local strains, we systematically collect near-field IR amplitude images with a spectral range from 1580 cm$^{-1}$ to 1590 cm$^{-1}$ for two representative regions- one with an individual wrinkle and the other one with a crack, as shown in Fig. 2a-d and 2f-i, respectively. For the region near the end of a wrinkle, our theoretical calculation (see Supporting Information) shows that there should exist local compressive strain, and consequently, the optical phonon frequency should blue shift to a value higher than the pristine phonon frequency of 1585 cm$^{-1}$, as illustrated in Fig. 2e. Whereas for the region near the end of a crack, there should exist local tensile strain, and the phonon frequency should have a red shift to a slightly lower frequency, as illustrated in Fig. 2j. Experimentally, at the end of the wrinkle, we observe a dark spot at lower excitation frequencies (Fig. 2a,b), which turns to a bright spot at higher frequencies (Fig. 2c,d). Such evolution of IR response can be well understood with the strain-induced change of phonon frequency and dielectric constant. As illustrated in Fig. 2e, the dielectric constant of the compressive strain region is smaller than that of unstrained pristine graphene at low frequencies (denoted by dashed lines a and b), while it is larger at higher frequencies (denoted by dashed lines c and d). At the end of the crack, near-field IR images (Fig. 2f-i) show opposite features: at lower frequencies, the local strain region appears as a bright spot; while at higher frequencies, it turns dark. The IR responses can in turn be well explained with a tensile strain as illustrated in Fig. 2j. Such agreement between experimentally measured near-field IR responses and theoretical analysis with a simple phonon-shift picture demonstrates unambiguously that the measured IR contrast is indeed induced by local strains.

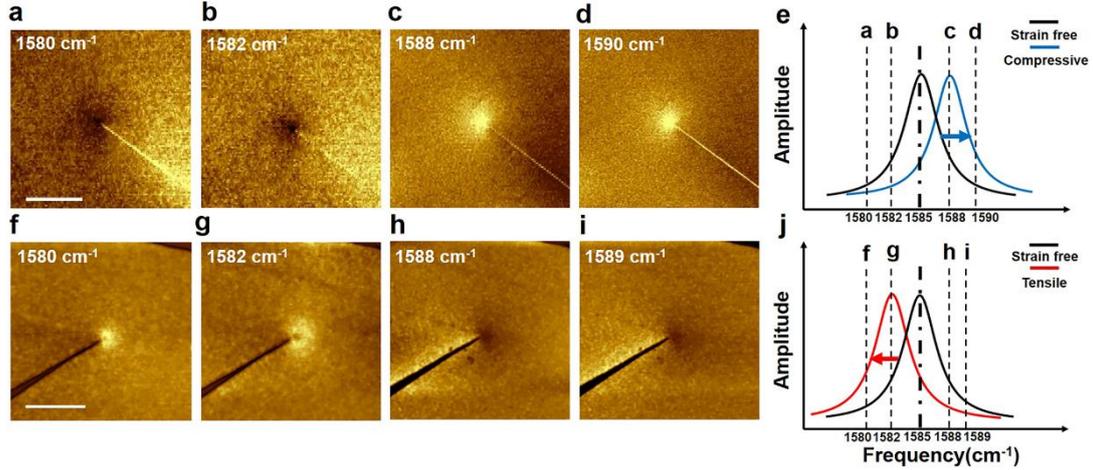

**Fig. 2. Frequency-dependent near-field IR response of both compressive and tensile strains.** (a-d) Near-field infrared nanoscopy images of the local strain region at the end of a wrinkle at different excitation frequencies. Theoretical analysis reveals this region to be compressive. In (a-b), at frequencies lower than intrinsic phonon frequency, near-field infrared nanoscopy images show a dark spot at the strained area. In (c-d), at frequencies higher than phonon frequency, near-field infrared nanoscopy images become a bright spot at the strained area. Scale bar 2 $\mu m$. (e) Illustration of compressive strain induced shift of phonon frequency and change of the dielectric function. The black and blue curves are characteristics of the dielectric function of strain-free and compressive-strained bilayer graphene. Dashed lines represent excitation frequencies as shown in (a-d). (f-i) Near-field infrared nanoscopy images of tensile local strain near the end of a crack at different excitation frequencies. In (f-g), at frequencies lower than phonon frequency, near-field infrared nanoscopy images feature a bright spot at the strained area. In (h-i), at frequencies higher than phonon frequency, near-field infrared nanoscopy images turn to a dark spot at the strained area. Scale bar 1 $\mu m$. (j) Illustration of tensile strain induced phonon frequency shift and dielectric function change. The black and red curves are characteristics of the dielectric function of strain-free and tensile-strained bilayer graphene. Dashed lines represent excitation frequencies as shown in (f-i).

We then demonstrate that electrical gating plays an important role in the detection of the local strain. Note that all the above IR amplitude images shown in Figs. 1 and 2 are obtained when graphene samples are highly doped. In order to show the importance of the electrical gating, we perform near-field IR measurement of bilayer graphene with and without electrical gating, for comparison. A series of infrared images taken at $V_g = 0$ V and $V_g = 60$ V in frequencies ranging from 1580 cm$^{-1}$ to 1590 cm$^{-1}$ are presented in Fig. 3. At $V_g = 0$ V (charge-neutral), wrinkles can be seen clearly in the IR images (Fig. 3a-d), but local strains near the wrinkle end can be hardly seen. An IR spectrum (Fig. 3e) of an unstrained position (denoted by the white arrows) extracted from a sequence of near-field IR images shows a largely flat feature with a rather weak phonon response. This is understandable as we mentioned that bilayer graphene is a nonpolar crystal and its phonon is infrared inactive. The observed tiny phonon response may be due to non-zero initial doping. After applying a gate voltage of 60 V, local strains can be clearly seen in the near-field IR images of Fig. 3f-i. This is because the graphene phonon is activated by the electrical gating. The phonon response is now largely enhanced and shows a prominent peak as shown in Fig. 3j.

A vertical gate electrical field breaks the symmetry of the two layers in bilayer graphene and dopes the two layers with unequal amounts of charge carriers. Optical phonons are now dressed by electron-hole pairs, and thus gain electrical dipole moment. Consequently, IR inactive phonons become IR active. Such activation of optical phonon by electrical gating enables the detection of phonon frequency as well as local strains at the nanometer scale with near-field IR spectroscopy. In addition, optical transitions generated by dressed phonon excitation interfere with continuous electronic transitions and form an asymmetric Fano resonance line shape. Such asymmetric Fano line shape holds an incredibly high slope (Fig. 3j), and can potentially boost the sensitivity of strain detection.

The observed near-field IR scattering $A(E)$ at a given excitation energy $E$ can be well fitted with the Fano formula[31]

$$A(E) = A_e * \frac{[q*\gamma+(E-E_0)]^2}{(E-E_0)^2+\gamma^2},$$

where $A_e$ is the bare electronic state scattering, $E_0$ and $\gamma$ are the center energy and linewidth of the phonon resonance, respectively. The dimensionless Fano parameter $q$ describes the ratio of the transition probabilities to the dressed phonon and the electronic states, and it determines the spectral shape. Note that the Fano formula applies not only to absorption but also to other different optical spectra, such as scattering and transmission. The Fano formula has been an important tool to design and analyze optical problems, because it can describe resonant phenomena in a variety of systems, including semiconductor nanostructures,[32] photonic crystals,[33] superconductors[34] and more others. Moreover, it can help to achieve negative optical scattering force for nanoparticles,[35] to reveal bound states in the continuum[36] and exotic states of subwavelength topological photonics,[37] as well as to realize a range of useful applications. To understand the physical picture of Fano resonance deeply, a detailed analysis of the Fano asymmetry parameter $q$, which is the key parameter of the Fano theory, must be provided. According to the value of $q$, the scattering line shape can be divided into resonance, dispersive, and anti-resonance. For the case of resonance, the Fano parameter $|q| \gg 1$, which means the phonon dominates. For the case of dispersive, parameter $|q| \approx 1$, which indicates the phonon and electron contribute equally. For the case of anti-resonance, parameter $|q| \ll 1$, which implies the electron dominates. The Fano parameter $q$ is extracted by fitting the scattering spectra with the Fano formula, and its value in Fig. 3e and 3j are 0 and 3.6, respectively. Parameter $q = 0$ means that there are only electronic transitions with no phonon contribution at $V_g = 0$ V. The observed small symmetric dip is unique to Fano resonance, sometimes called an antiresonance.[33,38] Parameter $q$ increases to 3.6 (larger than unit) at $V_g = 60$ V, implying that the transition to dressed phonon now dominates. It is interesting that in bilayer graphene the optical transitions to electrons and to phonons are tunable with electrical gating. This reveals bilayer graphene to be an exciting material for tunable nanophotonic and nanoelectronic devices and circuits in the mid-infrared range. Thanks to the tunability of the Fano resonance in bilayer graphene, we are able to detect graphene phonon frequency as well as local strains with near-field infrared nano spectroscopy.

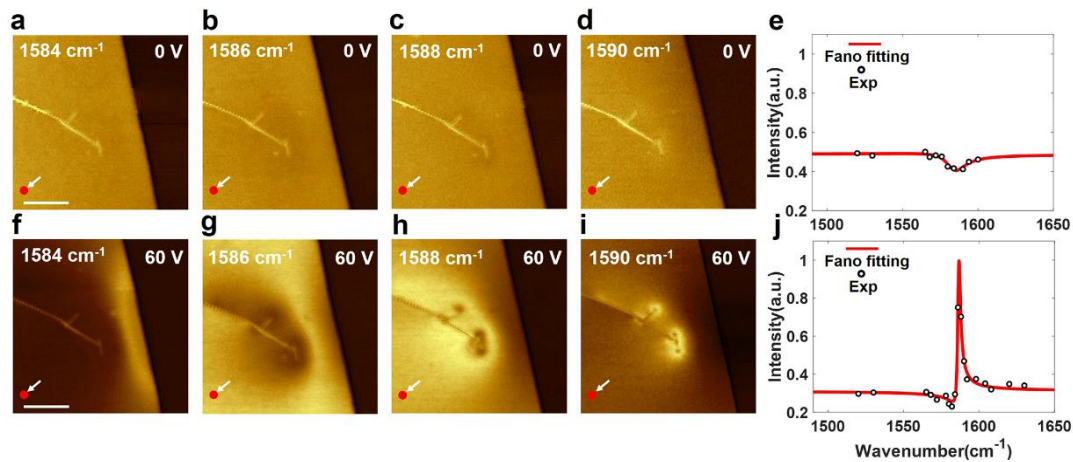

**Fig. 3. Nano-imaging of local strain enabled by Fano resonance through electrical gating.** (a-d) Infrared images of a bilayer graphene flake with wrinkles at different excitation frequencies, under a gate voltage of 0 V. Local strains around the wrinkles can be hardly seen. (e) IR spectrum taken at an unstrained position denoted by the white arrows in (a-d), displaying a rather weak phonon response. (f-i) Near-field infrared nanoscopy images of the same sample taken at a gate voltage of 60 V, from which strained region can be clearly identified. (j) IR spectrum taken at the same unstrained region as in (e) under a gate voltage of 60 V, showing a strong phonon response with Fano line shape. The vertical electrical field from electrical gating breaks the symmetry of the two graphene layers and introduces electrical dipole moment to graphene phonons. Such charge dressed phonons interact strongly with IR photons, featuring a Fano resonance. As a result, the detection of phonon frequency and local strain becomes possible. Scale bar 700 nm.

The quantitative phonon frequency shift of a strained area (Fig. 4a) is obtained from fitting near-field IR spectra and comparing it with that of intrinsic graphene. Five spectra are collected along the black dashed arrow (Fig. 4a) from an unstrained position (purple dot) to the center of a selected strained area (blue dot), and plotted in Fig. 4b. A systematic shift of phonon frequency can be clearly observed from those infrared spectra, and a shift of 0.5 cm$^{-1}$ of the strain center can be extracted. In addition, the peak width of the strained area also increases, which may be induced by the anisotropy of the strain field.

In order to obtain strain magnitude from the experimentally measured shift of phonon frequency, we calculate the relationship between strain and phonon frequency shifts. First-principles density function theory (DFT) calculations are performed within the local density approximation (LDA) to get the quantitative frequency shift of TO phonon mode in bilayer graphene under strain (see methods for details). The results can be summed as, with uniform in-plane strain, the TO phonons remain degenerate and the frequency shift linearly with strain; while with the high symmetry directions ($\Gamma$-M, $\Gamma$-K) anisotropic strain, the degenerate TO phonon resonance frequencies will break into two because of the breaking symmetry (inset of Fig. 4c). Moreover, both two TO phonon resonance frequencies shift linearly with the anisotropic strain. From the calculated results (Fig. 4c), the optical phonon resonance frequency of bilayer graphene will shift linearly by a slope of -57.34 cm$^{-1}$/% with a uniform in-plane strain along $\Gamma$ point to K point applied. As expected, the compressive strain will shift the phonon in bilayer graphene to a higher frequency. On the contrary, the tensile strain will shift the optical phonon to a lower frequency.

Based on the experimentally measured shift of phonon frequency of 0.5 cm$^{-1}$ and the theoretical shift slope of -57.34 cm$^{-1}$/%, we obtain the strain amplitude at the position denoted by the blue dot in Fig. 4a to be -0.01%. Strain amplitudes extracted from all the five IR spectra are plotted in Fig. 4d, which agree well with values calculated from pure geometrical analysis of the end of a wrinkle (see Supporting Information for details). Next, we briefly estimate the minimum detectable strain, i.e., the sensitivity of the detection technique. The strain is detectable only when the strain-induced change of the IR signal is greater than the noise level. On the one hand, the typical value of noise in our infrared images is ~ 2.5%. On the other hand, the phonon peak is very sharp due to Fano interference and the largest slope of the spectral lines in Fig. 4b is ~ 25%/cm$^{-1}$. Therefore, a frequency shift of 0.1 cm$^{-1}$ is the minimum detectable frequency shift. Considering the strain-induced frequency shift slope is ~ 50 cm$^{-1}$ /%, we therefore obtain the minimum strain that can be detected to be 0.002%, meaning that we are able to detect a relative lattice change of $2\times10^{-5}$ in bilayer graphene. Such sensitivity is higher than most strain detection techniques due to the Fano interference induced extremely sharp phonon peak.

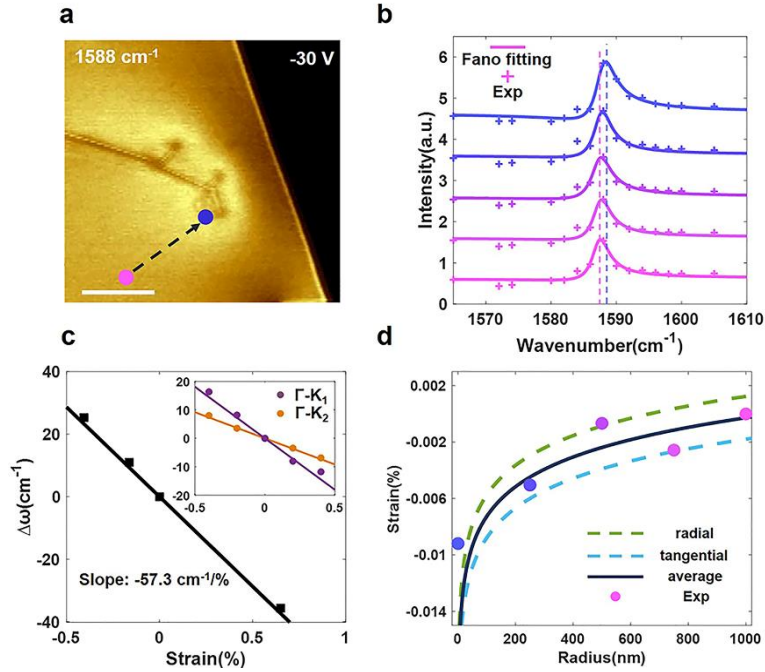

**Fig. 4. Quantitative analysis of local strains in gated-bilayer graphene.** (a) Near-field IR image of bilayer graphene under a gate voltage of -30 V. The excitation frequency is 1588 cm$^{-1}$. Scale bar 700 nm. (b) Spectra collected from pristine region to strained region along the black line displayed in (a), from which a gradual shift of resonance frequency can be observed. (c) First-principles calculation of frequency shift of optical phonon of bilayer graphene as a function of applied strain along direction Γ-K direction. (d) Theoretical calculations of local strain distribution in radial and tangential directions, and the average strain for the sample in (a). Experimental data (purple dots) are extracted from the phonon frequency shift in IR spectra in (b) and first-principles calculation results in (c).

In conclusion, we have demonstrated nanometer-scale detection of local strain in bilayer graphene through the mapping of optical phonon frequency with near-field IR spectroscopy. The probing of optical phonons in nonpolar bilayer graphene is enabled by dressing phonons with electron-hole pairs through electrical gating. The dressed phonon features a super sharp asymmetrical Fano lineshape, which boosts the strain detection sensitivity up to 0.002%. In addition, we have also presented a nanometer-scale tunable Fano system in bilayer graphene with electrical gating and SNOM probing. Our studies not only extend the application of SNOM in detecting phonons and strains in nonpolar crystals, but also provide exciting opportunities to study rich strain effects in

Bernal-stacked bilayer graphene, twisted bilayer graphene as well as ABC-stacked trilayer graphene.

**Methods**

**Near field IR imaging.** A homemade IR-SNOM is used for the near-field IR imaging, which includes a Bruker Innova AFM and a Daylight Solution Quantum Cascade laser (QCL). A mid-infrared light generated by the QCL laser is focused onto the apex of a metal-coated AFM tip. The enhanced optical field at the tip apex interacts with bilayer graphene underneath the tip. The backscattered light, carrying near-field optical information on the sample, is collected by the MCT detector (KLD-0.1-J1, Kolmar) placed in the far field. Near-field scattering amplitude is extracted through a homodyne detection scheme. Near-field optical images with spatial resolution better than 20 nm can be achieved with sharp AFM tips. Near-field IR images can be recorded simultaneously with the topography information during the experiment.

**DFT calculations.** The first-principles density functional theory (DFT)[39,40] calculations are carried out by the projector augmented wave (PAW)[41] method with the local-density approximation (LDA)[42] as implemented in the VASP package[43]. The experimental lattice constant a = 2.46 Å together with the inter-layer-distance c = 3.35 Å is used in the bilayer graphene calculations, and a 15 Å vacuum layer is added to minimize the interactions between bilayers in the neighboring supercells along the z direction. The phonon frequency is calculated within the PHONOPY package[44] by using a $4 \times 4 \times 1$ supercell, with a $40 \times 40 \times 1$ k-mesh and a plane-wave cutoff of 500 eV.

**Supporting Information**

(1) Near-field IR nano-imaging of compressive strain in bilayer graphene, (2) Near-field IR nano-imaging of tensile strain in bilayer graphene, (3) First-principles calculations of strain induced phonon frequency shift in bilayer graphene.

**Acknowledgements**

This work is supported by the National Key Research and Development Program of China (2016YFA0302001) and the National Natural Science Foundation of China (11774224, 12074244,

11521404 and 61701394). Z.S. acknowledges support from the Program for Professor of Special Appointment (Eastern Scholar) at Shanghai Institutions of Higher Learning, and additional support from a Shanghai talent program. We would also like to thank the Centre for Advanced Electronic Materials and Devices (AEMD) of Shanghai Jiao Tong University (SJTU) for the support in device fabrication. First-principles DFT calculations were performed at the Center for High Performance Computing of Shanghai Jiao Tong University.